\documentclass[aps,prl,reprint,superscriptaddress,floatfix,preprintnumbers]{revtex4-1}
\usepackage{graphicx}
\usepackage{textcomp}
\usepackage{hyperref}
\usepackage{amsmath}
\usepackage{amsfonts}
\usepackage{color}

\hypersetup{
    colorlinks=true,
    linkcolor=black,
    citecolor=black,
    filecolor=black,
    urlcolor=black,
}

\begin{document}


\title{Shape-morphing architected sheets with non-periodic cut patterns}


\author{Paolo Celli}
\email{pcelli@caltech.edu}
\affiliation{Department of Mechanical and Civil Engineering, California Institute of Technology, Pasadena, CA 91125, USA}
\author{Connor McMahan}
\affiliation{Department of Mechanical and Civil Engineering, California Institute of Technology, Pasadena, CA 91125, USA}
\author{Brian Ramirez}
\affiliation{Department of Mechanical and Civil Engineering, California Institute of Technology, Pasadena, CA 91125, USA}
\author{Anton Bauhofer}
\affiliation{Department of Mechanical and Civil Engineering, California Institute of Technology, Pasadena, CA 91125, USA}
\affiliation{Department of Mechanical and Process Engineering, ETH Zurich, 8092 Zurich, Switzerland}
\author{Christina Naify}
\affiliation{Jet Propulsion Laboratory/California Institute of Technology, Pasadena, CA, 91109, USA}
\author{Douglas Hofmann}
\affiliation{Jet Propulsion Laboratory/California Institute of Technology, Pasadena, CA, 91109, USA}
\author{Basile Audoly}
\affiliation{Department of Mechanical and Civil Engineering, California Institute of Technology, Pasadena, CA 91125, USA}
\affiliation{Laboratoire de M\'{e}canique des Solides, CNRS, UMR 7649, D\'{e}partement de M\'{e}canique, \'{E}cole Polytechnique, 91128 Palaiseau CEDEX, France}
\author{Chiara Daraio}
\email{daraio@caltech.edu}
\affiliation{Department of Mechanical and Civil Engineering, California Institute of Technology, Pasadena, CA 91125, USA}


\begin{abstract}
\vspace{5px}
\noindent{\textbf{This article may be downloaded for personal use only. Any other use requires prior permission of the authors and RSC Publishing. This article appeared in}: \emph{Soft Matter} {\bf 14}(11), 9744--9749 (2018) \textbf{and may be found at}: \url{https://doi.org/10.1039/C8SM02082E}}
\vspace{15px}

We investigate the out-of-plane shape morphing capability of single-material elastic sheets with architected cut patterns that result in arrays of tiles connected by flexible hinges. We demonstrate that a non-periodic cut pattern can cause a sheet to buckle into three-dimensional shapes, such as domes or patterns of wrinkles, when pulled at specific boundary points. These global buckling modes are observed in experiments and rationalized by an in-plane kinematic analysis that highlights the role of the geometric frustration arising from non-periodicity. The study focuses on elastic sheets, and is later extended to elastic-plastic materials to achieve shape retention. Our work illustrates a scalable route towards the fabrication of three-dimensional objects with nonzero Gaussian curvature from initially-flat sheets.
\end{abstract}


\maketitle


Imparting elastic sheets with a mesoscale architecture enables the creation of materials with unusual characteristics, such as extreme extensibility~\cite{Tang_EML_2017}, deployability~\cite{Filipov_PNAS_2015, Overvelde_NATURE_2017} and auxeticity~\cite{Grima_PSS_2005, Mullin_PRL_2007}. These properties can be leveraged to design sheets that morph into complex three-dimensional objects. For example, origami sheets can be turned into nearly-arbitrary shapes~\cite{Schenk_PNAS_2013, Silverberg_SCIENCE_2014, Dudte_NATMAT_2016, Callens_MATERTODAY_2017}, but are typically challenging to fold~\cite{Demaine_SOCG_2017} or actuate~\cite{Hawkes_PNAS_2010, Stern_PRX_2017, Plucinsky_SOFTMATTER_2018}. Patterned elastomeric sheets~\cite{Klein_SCIENCE_2007, Kim_SCIENCE_2012, Ge_APL_2013, Ware_SCIENCE_2015, Gladman_NATMAT_2016, Aharoni_PNAS_2018}, bilayers~\cite{vanRees_PNAS_2017} and sheets with smart hinges~\cite{Felton_SOFT_2013, NA_ADVMAT_2014, Ahmed_SMS_2014} can morph into three-dimensional surfaces with nonzero Gaussian curvature via non-mechanical stimuli, but their fabrication is complex. Ribbon- and membrane-like flat mesostructures can buckle out of plane and produce three-dimensional geometries when subject to mechanical actuation~\cite{Xu_SCIENCE_2015, Zhang_PNAS_2015, Dias_SOFTMATTER_2017, Guo_NPJFLEX_2018}. However, compressive actuation requires non-trivial assembly processes, and the geometries obtained via tensile loads are limited to thin, arch-like features.

In contrast to shape-morphing origami or bilayer films, sheets with architected cut patterns can be easily fabricated via subtractive technologies. Their out-of-plane deformation can be triggered by manual forming~\cite{Sussman_PNAS_2015, Konakovic_ACM_2016, Wang_JAM_2017}, via the actuation of smart hinges~\cite{Cui_SMS_2017}, or by applying compressive boundary loads ~\cite{Zhang_PNAS_2015, Neville_SCIREP_2016, Fu_NATMAT_2018}. Recently, it has been demonstrated that sheets with periodic perforations can also buckle locally in tension~\cite{Blees_NATURE_2015, Shyu_NATMAT_2015, Rafsanjani_PRL_2017, Tang_ADMA_2017, Guo_NPJFLEX_2018}, producing crease patterns that can be used for soft robotic locomotion~\cite{Rafsanjani_SCIROB_2018} or as coatings for sunlight control~\cite{Tang_ADMA_2017}. However, since these buckling modes take place at the scale of the unit cells, the size of the transverse features they can produce cannot significantly exceed the typical length of the cuts. Non-periodic cut patterns have been seldom explored in this context: non-periodicity is known to lead to geometric frustration~\cite{Coulais_NATURE_2016, Guseinov_ACM_2017}, i.e., the desired deformation mode is impeded by the geometric incompatibility between neighboring cells. In the few cases where non-periodic cut patterns have been explored, frustration has been avoided~\cite{Cho_PNAS_2014, Grima_ADVMAT_2016, Ion_UIST_2016, Mirzaali_SCIREP_2018}. In particular, the effect of geometric frustration on the out-of-plane deformations of thin architected sheets has been ignored so far.

In this work, we study the tensile response of elastic sheets featuring non-periodic cut patterns, and intentionally leverage geometric frustration to induce controllable, global shape changes via buckling. In most of our designs, we use point-like boundary loads that induce large deformations in selected sub-domains of the sheets. The inhomogeneous distribution of strains results in global buckling modes that make the sheets bend out of plane and morph into dome-like surfaces with nonzero Gaussian curvature and patterns of wrinkles confined to pre-determined regions. Note that a similar mechanism is at work in the morphogenesis of living systems due to differential growth~\cite{Efrati_JMPS_2009, Boudaoud_TRENDSPLANTSCIENCE_2010, Osterfield_DEVCELL_2013}. We also extend the method to initially cylindrical sheets and to cut patterns arranged into non-rectangular grids, and demonstrate the formation of persistent three-dimensional surfaces via elastic-plastic materials~\cite{Rafsanjani_PRL_2017}. Our work distinguishes itself for the simplicity of fabrication and actuation, and for its potential applicability to material and structural systems at vastly-different scales; it therefore illustrates a potential path towards the high-throughput realization of morphable surfaces. 

\begin{figure} [!htb]
\centering
\includegraphics[scale=1.40]{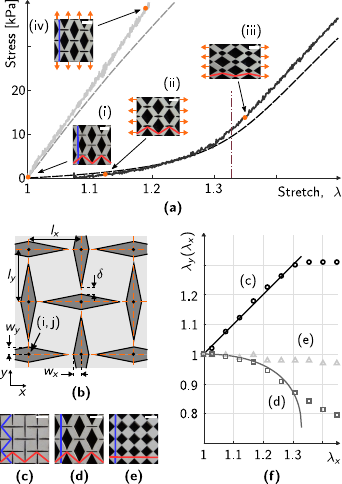}
    \caption{ In-plane response of periodic sheets. (a) Uniaxial response of the sheet with undeformed geometry shown in~(i): horizontal loading (black lines) and vertical loading (grey lines), experiments (solid lines) and FE simulations (dashed lines). The vertical dash-dot line is the geometric-to-elastic transition predicted by kinematics. Insets (i-iv) show snapshots of a $4\times4$-tile portion of the sheet at different levels of deformation. The red and blue overlaid lines are obtained by joining the diagonals in a particular row and column of tiles. (b) Sketch of a generic periodic architecture parameterized by design variables. (c-e) Details of three periodic undeformed specimens, corresponding to different values of $(w_x, w_y)$ listed in the SI. (f) Transverse stretch $\lambda_y$ as a function of the stretch along the loading direction, $\lambda_x$, for the same set of specimens: experiments (markers) and kinematic predictions (solid curves). Scale bar: 6 mm}
\label{fig:tensile}
\end{figure}

We start by analyzing a simple cut pattern featuring a large-amplitude, planar mode of deformation. A $108$-by-$108\,\mathrm{mm}$, $1.55\,\mathrm{mm}$-thick natural rubber sheet is laser-cut~\cite{Rafsanjani_EML_2016} following a pattern of diamond-shaped cut-outs and straight cut lines ending close to the diamonds' vertices. These two types of cuts are visible in black in the insets of Fig.~\ref{fig:tensile}. 
The result is an array of 18$\times$18 rhomboid tiles connected by thin hinges. The experimental traction curves (Fig.~\ref{fig:tensile}) for unaxial tension reveal a strongly anisotropic and non-linear behavior (see SI for details). When the tension is applied in the $x$-direction (horizontal direction in the figure, black lines) the response is initially compliant up to a stretch value $\lambda \sim 1.3$, and then becomes stiffer. When the tension is applied in the $y$-direction (vertical direction in the figure, gray lines), the response is stiff and approximately linear, without any compliant regime. We simulated the mechanical response of the architected sheet numerically, by using a finite element (FE) model for a neo-Hookean material in plane strain (an assumption justified in the \emph{Supplementary Information}, SI). The numerical traction curves are in good agreement with the experimental ones (Fig.~\ref{fig:tensile}).

The salient features of the loading curves can be explained by a kinematic analysis, in which the sheet is modeled as an array of rigid tiles connected by pin joints. Such networks can feature modes of deformation known as \emph{mechanisms}~{\cite{Pellegrino_IJSS_1986}}, which are mapped to low-energy configurations of the elastic sheet involving mainly bending and shear at the hinges~\cite{Coulais_NATPHYS_2017}. A mechanism relies on the coordinated rotation of the tiles in response to applied tension (Fig.~\ref{fig:tensile}, SI and~\cite{Hutchinson_JMPS_2006, Kapko_PRSA_2009}). The maximum stretch attainable via a mechanism can be derived by considering the broken lines connecting the diagonals of the tiles in a given row or column---red and blue lines in Fig.~\ref{fig:tensile}(i). As the length of these lines is preserved by mechanisms, the maximum stretch in the $x$ or $y$ direction is attained when the corresponding line is fully stretched out. For the cut design used in Fig.~\ref{fig:tensile}, this maximum stretch is calculated by a geometric argument as $\lambda_x = 1.33$ in the $x$-direction, as indicated by the dash-dotted line in the figure; this is indeed where the compliant-to-stiff transition is observed in the traction curves. Conversely, the (blue) line of diagonals in the $y$-direction is straight by design, and no mechanism can be activated when the tension is applied in this direction; this is consistent with the absence of an initial compliant regime in the grey curves in Fig.~\ref{fig:tensile}.

Next, we introduce a family of periodic cut patterns parameterized by design variables. Our generic pattern, sketched in Fig.~\ref{fig:tensile}(b), is obtained by cutting out diamonds with alternating directions, centered at the nodes of a grid of $N_x \times N_y$ rectangles, each with dimensions $l_x \times l_y$. The two families of diamonds are assigned different widths, $w_x$ and $w_y$, so that the previous design comprising line-cuts can be recovered as the special case $w_y = 0$. The length of the diamonds is such that a gap (hinge) of width $\delta$ is present between adjacent diamonds. Three examples of periodic geometries cut out in natural rubber sheets are shown in Fig.~\ref{fig:tensile}(c-e), for $N_x = N_y = 18$ and $l_x=l_y=6\,\mathrm{mm}$; note that the shape of the tiles (light grey) can now vary from rhomboid to square. Experimental traction curves for three particular cutting patterns are plotted in the plane of stretches $(\lambda_x, \lambda_y)$ in Fig.~\ref{fig:tensile}(f), and compared with the predictions of the kinematic analysis (see SI),
\begin{equation}
\lambda_y\left(\lambda_x\right)=\frac{d_v}{l_y}\,\sin\!\left[\gamma+\arccos\left( \frac{\lambda_x l_x}{d_h} \right)\right]\,,
\label{eq:kinematicModel}
\end{equation}
where $d_h$ and $d_v$ are the lengths of the diagonals of a tile, and $\gamma$ is the angle between these diagonals. The design variables have a strong influence on tension tests. The cut pattern in Fig.~\ref{fig:tensile}(c) gives rise to an auxetic mechanism~{\cite{Grima_PSS_2005}} having a negative Poisson's ratio $\nu = -1$; this is reflected by the positive slope of the black curve in Fig.~\ref{fig:tensile}(e). By contrast, the mechanism associated with the cut pattern in Fig.~\ref{fig:tensile}a,d has a positive Poisson's ratio. For both these cut patterns, the kinematic model  provides an accurate prediction of the transverse stretch up to around $\lambda_x \sim 1.3$, where the joints start to stretch. Finally, the cut pattern in Fig.~\ref{fig:tensile}(e) is stiff when loaded in tension since the diagonals of adjacent tiles are aligned. The effect of the design parameters $\delta$ and $t$ on the in-plane response is discussed in the SI.

\begin{figure} [!htb]
\centering
\includegraphics[scale=1.40]{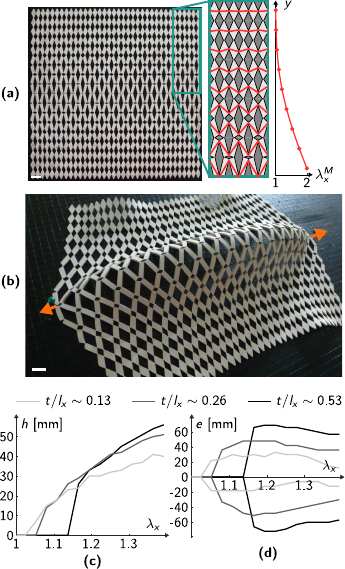}
  \caption{Out-of-plane morphing of a graded sheet. (a) Cut pattern with gradient in the $y$-direction. The inset to the right highlights the tile diagonals (in red), and the corresponding maximum kinematic stretch $\lambda_x^M (y)$: the sheet is highly stretchable at the center, but inextensible away from it. (b)~A dome shape obtained when the sheet is pulled from two boundary points, as indicated by the arrows. (c) Maximum height and (d) lateral extent of the dome for graded specimens with different thicknesses, for various stretches. Details on how these quantities were measured are given in the SI.}
\label{fig:graded}
\end{figure}

Having analyzed a family of \emph{periodic} cut patterns, we now investigate \emph{non-uniform} designs, obtained by specifying values of $w_x$ and $w_y$ in every cell of a rectangular grid; the cell size $l_x \times l_y$ is uniform throughout the sheet. Upon deformation, we expect that every unit cell of these non-periodic sheets will try to follow the mechanism corresponding to the local values of $w_x$ and $w_y$, as described by Equation~(\ref{eq:kinematicModel}). However, mechanisms corresponding to neighboring cells are not geometrically compatible in general (see SI). Thus, we investigate how this incompatibility is resolved at the global level by buckling. As a first example, we consider a cut geometry where $w_x$ is constant while $w_y$ varies sinusoidally in the $y$-direction, see Fig.~\ref{fig:graded}(a), 
using a $1.55\,\mathrm{mm}$-thick natural rubber sheet with $N_x = 36$, $N_y = 18$, $l_x = 6\,\mathrm{mm}$, $l_y = 2\,l_x$. This choice of maps for $w_x$ and $w_y$ ensures that the top and bottom parts of the sheet are virtually undeformable, see the inset in Fig.~\ref{fig:graded}(a), while the center is highly stretchable. When the sheet is stretched by point-like forces, as in Fig.~\ref{fig:graded}(b), the strong geometric incompatibility between the center and the edges produces a global buckling mode spanning the central region and featuring nonzero Gaussian curvature. Note that this buckling instability takes place in tension, unlike in the classical Euler buckling. Increasing the sheet's thickness $t$, we increase its effective bending modulus and the onset of buckling occurs at larger stretches, as shown in Fig.~\ref{fig:graded}(c). An increased thickness also yields larger deflections and makes the buckled pattern wider, as shown in Fig.~\ref{fig:graded}(d), where we report the lateral extent of the dome versus $\lambda_x$. These results, further discussed in the SI, illustrate that $t$ and $\lambda_x$ offer some control over the buckled shape and curvature. It is worth pointing out that to obtain buckled patterns as in Fig.~\ref{fig:graded} it is sufficient to have a variation of stretchability along a single axis (with or without auxeticity), {and to apply localized boundary loads along the direction of maximum stretch}.

\begin{figure} [!htb]
\centering
\includegraphics[scale=1.40]{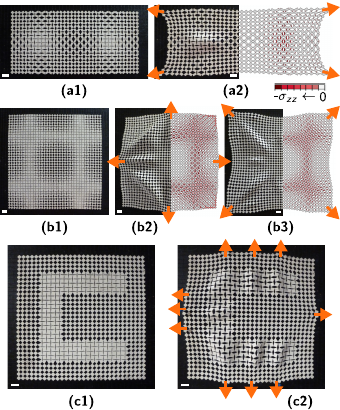}
  \caption{More complex cut patterns. (b1)~Specimen featuring two soft and auxetic regions in its interior, which give rise to two localized bumps upon pulling at the four corners (b2). (c1-c3) Response of another specimen, highlighting the influence of the boundary loading on the surface morphology. The right-halves of (b2), (c2) and (c3) are the stress maps of $\sigma_{zz} = \nu (\sigma_{xx} + \sigma_{yy})$ (under the plane strain assumption); negative values  are taken as an indicator for buckling. (d1-d2)~Shaping wrinkles: a C-shaped soft and auxetic region is embedded in a sheet by a suitable choice of the maps of $w_x$ and $w_y$ in the reference configuration (d1). The wrinkles localize upon the application of boundary loads (d2). The arrows indicate the boundary loads. Scale bar: 12 mm.}
\label{fig:nonperiodic}
\end{figure}

More complex buckling patterns can be obtained by letting both $w_x$ and $w_y$ vary along the sheet, either smoothly or abruptly. In these cases, a sufficient condition to obtain buckled surfaces is the presence of auxetic islands surrounded by unstretchable and non-auxetic regions. As an example, we study the sheet in Fig.~\ref{fig:nonperiodic}(a1); 
when stretched as indicated by the arrows, the two auxetic islands tend to swell biaxially, resulting in strong geometric incompatibilities. This swelling is prevented by the surrounding stiff regions, compressive in-plane stresses arise, and two domes localized on the auxetic islands appear; this is shown numerically and experimentally in Fig.~\ref{fig:nonperiodic}(a2). {Note that, for this particular cut pattern, a similar buckled shape can be obtained by replacing point loads with distributed boundary loads (see SI).} As another example, we study the response of a sheet with a more complex cut pattern obtained by varying both $w_x$ and $w_y$ sinusoidally along both the horizontal and vertical directions. The experimental results in (b2) and (b3), corresponding to actuation at the structures' corners or boundary mid-points, respectively, show markedly different wrinkle patterns, thereby highlighting the role of the applied force in selecting the pattern. Finally, in Fig.~\ref{fig:nonperiodic}(c1-c2), we show the response of a sheet featuring a C-shaped auxetic region inserted into an unstretchable sheet. In this case, pulling the specimen as indicated by the orange arrows leads to wrinkles localized along the C-like domain. The wavelength of the wrinkles is comparable to the width of the C-shaped domain. These examples show that the buckling patterns can be tailored by engineering the sheet's local properties through the maps of $w_x$ and $w_y$, and by choosing the points of application of the load.

\begin{figure}  [!htb]
\centering
\includegraphics[scale=1.40]{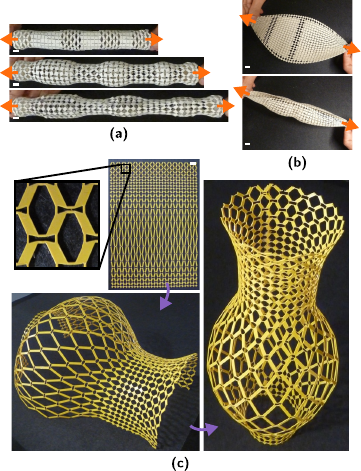}
  \caption{ (a) An architected tube can expand or contract radially based on an initial stripe pattern. Beyond a critical tensile load, an azimuthal buckling pattern appears in the expanded regions. (b) Petal-shaped specimen generated from a non-rectangular grid. This sheet morphs into a pea pod-shaped object when pulled from its ends. (c)~Sculpting axisymmetric shapes from a sheet made of an elastic-plastic material; the shapes are obtained by using graded cut patterns and by stretching out the sheets locally by hand. Scale bars: 12 mm.}
\label{fig:last}
\end{figure}

Similar principles can be extended to solids of revolution. For example, we pattern a sheet by varying $w_x$ and $w_y$ in vertical stripes, alternating regions of auxetic and not-auxetic behavior. We then roll the sheet, forming a tube, and pull on its ends. The applied tractions force the tube to expand at prescribed, auxetic sections and to contract at others~{\cite{Liu_SOFTMATTER_2018}}, see Fig.~\ref{fig:last}(a).
Stretching the tube further produces a non-axisymmetric buckling bifurcation, with an azimuthal wavelength roughly comparable to the stripes' width. Cut patterns can also be attached to non-Cartesian grids, as illustrated in Fig.~\ref{fig:last}(b), where the petal-like sheet closes up into a pea pod shape when pulled at its ends.

Finally, permanent three-dimensional shapes can be obtained by using an elastic-plastic material~\cite{Rafsanjani_PRL_2017}. This requires modifying the hinge design to avoid breakage: the new design, shown in the insets in Fig.~\ref{fig:last}(c), was inspired by~{\cite{Shang_JMR_2018}} (see SI). 
We leverage the elastic-plastic behavior to sculpt axisymmetric shapes out of a planar PETG sheet, as in Fig.~\ref{fig:last}(c). As earlier with the tube, this cut pattern is graded along the axis, which allows us to prescribe the radial expansion as a function of the axial coordinate. To obtain an even larger stretchability contrast, we use a non-regular rectangular grid, by setting $l_y(y)$ to take on larger values in the regions of large stretch. The irreversible deformations are obtained by stretching the sheet locally by hand, and a similar effect could be achieved using localized smart-material actuators or pressurized membranes. These structures are reminiscent of gridshells~\cite{Baek_PNAS_2018} and are easier to fabricate, especially at small scales.

In this work, we have demonstrated that geometric incompatibility can be leveraged to create three-dimensional objects from sheets with non-periodic cut-outs. By choosing the properties of the cuts locally, one can prescribe a map of maximum stretch, which is resolved when the sheet deforms out of plane in response to boundary loads or local stretching. 
While the shapes we have obtained are relatively simple, similar principles could be extended to different families of mechanisms, and could be coupled to optimization and inverse-design strategies to obtain more complex shapes. Towards this goal, it will be necessary to develop efficient numerical models of these systems and of their buckling behavior---a goal most likely achievable via homogenization. In conclusion, our work expands the shape-morphing gamut of perforated mono-layer sheets, and it indicates an approach that could be used to produce morphing and deployable structures at vastly-different scales. 

\begin{acknowledgments}
This research was carried out at the California Institute of Technology and the Jet Propulsion Laboratory under a contract with the National Aeronautics and Space Administration, and funded through the President's and Director's Fund Program. This work is partially supported through the Foster and Coco Stanback Space Innovation Fund. P.C. wishes to thank D. Pasini of McGill University, A. Constantinescu of \'{E}cole Polytechnique, and members of C.D.'s research group for their input and suggestions. We also thank B. Dominguez of Caltech for his assistance during laser cutting.
\end{acknowledgments}


%

\clearpage
\widetext

\setcounter{figure}{0}
\setcounter{page}{1}
\setcounter{section}{0}
\renewcommand{\thefigure}{S\arabic{figure}}
\renewcommand{\theequation}{S\arabic{equation}}
\renewcommand{\thesection}{S\arabic{section}}
\renewcommand{\thepage}{S\arabic{page}}
\makeatletter

\section*{Supporting Information for ``Shape-morphing architected sheets with non-periodic cut patterns''}

\section{Specimen fabrication}
\label{sec:fab}
A Universal ILS9 120W laser cutter is used to create perforations. We mainly use $1.55\,$mm-thick natural rubber sheets (McMaster-Carr, item no. 8633K71), but some $3.1\,$mm- and $0.75\,$mm-thick ones were also used (Grainger, items no. 1XWE5 and 8611K18). For the $1.55\,$mm-thick specimens, the machine is set to cut at 35\% power and 5\% speed, with an air assist flow rate of 100\% to avoid burning the specimens. For the $3.1\,$mm-thick specimens, 45\% power and 2.3\% speed are selected. For the $0.75\,$mm-thick specimens, 30\% power and 5\% speed are selected. Since the laser beam has a finite cutting diameter, the hinges are not characterized by sharp corners but are de-facto beams having a finite length. The tube specimens are closed using double-sided tape glued to some tiles. PETG sheets ($0.5\,$mm-thick) were perforated with the same laser cutter, with 3.0\% power and 2.2\% speed, and were also closed into surfaces of revolution using double-sided tape.

\section{Additional information on the tensile tests}
\label{sec:exp}
\begin{figure*} [!htb]
\centering
\includegraphics[scale=1.40]{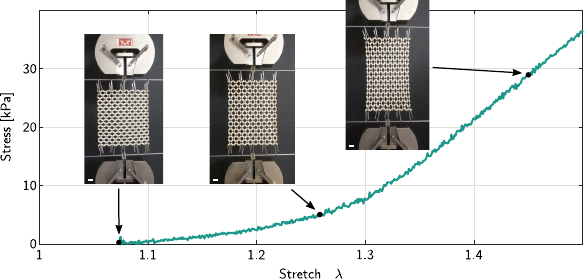}
  \caption{ Response of a $18\times18$ tile, anisotropic sheet with $\delta/l_x=1/8$ and $t/l_x\sim 0.26$. The insets depict the experimental setup and the response at three stretch values. (Scale bar, 12 mm)}
\label{fig:tensile2}
\end{figure*}

Uniaxial tensile tests are conducted using an Instron ElectroPuls (Model E3000) system equipped with a 250 N load cell at a constant deformation rate of $2\,\mathrm{mm\,s^{-1}}$. The tensile forces and displacements are measured with $1\,\mathrm{mN}$ and $5\,\mathrm{\mu m}$ accuracy, respectively, at an acquisition rate of 1 kHz. The force-displacement data obtained from the Instron WaveMatrix software is converted to stress-stretch data using the original sample dimensions. The data obtained is then subsampled to remove some of the noise (one every 10 measurements is kept). The response of the specimen in Fig.~\ref{fig:tensile}(a) is replicated in Fig.~\ref{fig:tensile2}, where the insets represent a few stages of the deformation of the specimen and show the experimental setup.
To accomodate lateral expansions and/or contractions of the specimens undergoing tensile loads, we employ a fixture where specimens are hung in a curtain-like fashion. We use 3D-printed parts (Formlabs Form 2, clear resin) to connect horizontal steel rods to the Instron's clamps; we then use paper clips as ``hooks'' to hang the specimens (at 5 locations on each side). Upon pulling, the paper clips can slide on the steel rod; the friction between these components will inevitably affect the response. Note that, due to the very small forces involved in our experiments, we claim that the elasticity of paper clips and steel rods only minimally affects the response. From Fig.~\ref{fig:tensile2}, we see that the response is recorded only for values of stretches larger than $\sim 1.08$. This is due to the fact that, when attached to our fixtures, some of the sheets we consider tend to deform due to their self weight. This self-stretching happens only when specimens feature mechanism-like deformation in the pulling direction. For example, in Fig.~\ref{fig:tensile}(a), the curve corresponding to horizontal stretching starts at 1.08, while the one for vertical stretching starts at 0.

\begin{figure*} [!htb]
\centering
\includegraphics[scale=1.40]{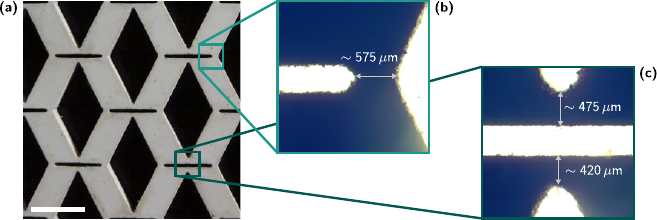}
  \caption{ (a) Detail of one of the anisotropic architectures analyzed in this work (Scale bar, 6 mm). (b) and (c) Microscope images (2.5$\times$ zoom) representing the details of vertical and horizontal hinges, respectively.}
\label{fig:anisotropicmicro}
\end{figure*}

From Fig.~\ref{fig:tensile}(a), we can see that the slopes of the elasticity-dominated portions of the experimental curves corresponding to horizontal and vertical stretching are not identical. This is caused by the fact that the size of the vertical and horizontal hinges in our anisotropic specimens are not identical. This is clearly visible from Fig.~\ref{fig:anisotropicmicro}.
In particular, the laser cutting process causes vertical hinges to be thicker than the horizontal ones. This explains why in Fig.~\ref{fig:tensile}(a) the continuous light gray curve is steeper than the elasticity-dominated portion of the continuous black curve.

In Fig.~\ref{fig:tensile4} we report the tensile response of the isotropic auxetic architecture displayed in Fig.~\ref{fig:tensile}(c) and Fig.~\ref{fig:mechanism}(a).
\begin{figure} [!htb]
\centering
\includegraphics[scale=1.40]{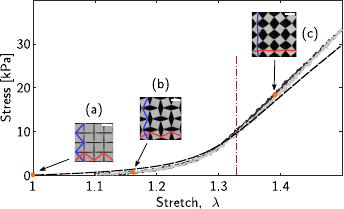}
  \caption{Tensile response of a periodic sheet featuring the undeformed architecture in (a). Black lines represent the sheet's response to horizontal stretching and light gray lines to vertical stretching. Solid lines are experimental curves. The dashed line represent the numerical response to both horizontal and vertical loading. The vertical dash-dot line shows the theoretically-predicted value for the transition from a mechanism-dominated deformation to an elastic deformation. Insets (a-c) show different stages of the sheet's deformation (Scale bar, 6 mm); the red and blue lines highlight the diagonals of each tile in a given row and column, respectively.}
\label{fig:tensile3}
\end{figure}
The two continuous lines, dark and light, represent the experimental curves obtained by pulling the specimen along the horizontal and vertical directions, respectively. The two almost overlap, as expected, due to the isotropic nature of the specimen's response. The dashed line is obtained from FE simulations. The superimposed dash-dot curve represents the analytical mechanism-to-elasticity transition.

\clearpage
\section{Details on the finite element model}
\label{sec:fem}
\begin{figure} [!htb]
\centering
\includegraphics[scale=1.40]{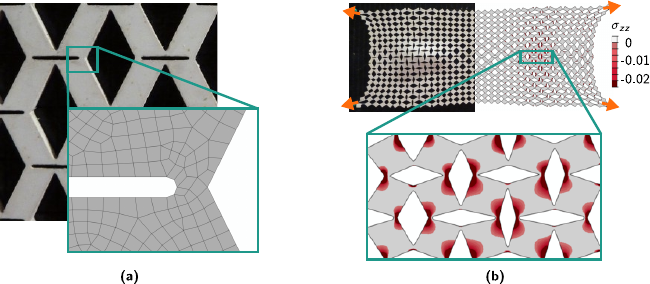}
  \caption{ Details of the FE model. (a) Detail of the mesh used for one of the hinges in the simulation of the anisotropic specimen tensile test. (b) Detail of the stress map for the simulation in Fig.~3(a2).}
\label{fig:fem}
\end{figure}

In this work, finite element simulations are carried out using Abaqus/Standard. The investigated sheets present different lengthscales: the hinge in-plane width and length ($\sim 1\,$mm), the length of a tile ($\sim 10\,$mm), and the total size of the sheet ($\sim 100\,$mm). Since the mechanical behavior of the sheets is, to a large extent, governed by the design of the hinges, a sufficiently fine mesh is required to accurately capture the correct response. Another challenge stems from the large nonlinearities involved and from the large distortions happening at large stretches. In order to efficiently identify regions that are prone to out-of-plane bending, we conduct two-dimensional finite element simulations. In all simulations, we resort to a plane strain assumption, accounting for the fact that the response is primarily determined by the hinge dimensions, and the hinges' in-plane width ($\sim 0.5\,$mm) is smaller than their out-of-plane thickness ($\sim 1.55\,$mm). Throughout this work, we consider geometric nonlinearities and model the nonlinear material behavior of natural rubber gum with a Neo-Hookean material model. This model is fit to the experimental response of a natural rubber dogbone specimen to tensile loading. Fig.~\ref{fig:fem}(a) shows a detail of the mesh at one of the hinges. 
We check mesh convergence for one of the simulations used to obtain the numerical curves in Fig.~\ref{fig:tensile}(a). We change the element size and monitor the stress values recorded for a given stretch along a given direction. The errors we obtain for doubling the average element size are below 0.73\%.

The results reported in Fig.~\ref{fig:tensile}(a) and Fig.~\ref{fig:tensile3} show that the numerics capture the features observed experimentally, even though some discrepancies between experiments and numerics arise at large stretches. These discrepancies can be attributed to several factors: 1) the inability of the Neo-Hookean model to capture the correct mechanical behavior at large stretches; 2) the fact that the CAD models used for our numerical simulations do not account for the exact hinge dimension that results from the laser cutting process; 3) the simulated loads might not be exactly identical to the experimental ones.

The stress maps in Fig.~3 represent the out-of-plane stress $\sigma_z=\nu(\sigma_x+\sigma_y)$. The colormap is designed to give relevance only to compressive stresses---those that are responsible for the onset of buckling. The stresses are not averaged over subdomains. Thus, the red areas in Fig.~\ref{fig:fem}(b) correspond to the compressive stress of the hinges. We also observe that the compressive stresses partially percolate into the tiles. This is likely responsible for out-of-plane buckling. From Fig.~3(b2-b3), we can see that the stress maps for the two loading configurations are almost identical. For this reason, the stress maps do not contain enough information to determine the exact shape of the resulting buckling patterns in complex scenarios, but give a useful guideline on where buckling is likely to occur in simple cases like that depicted in Fig.~3(a1-a2).

\clearpage
\section{Kinematic analysis}
\label{sec:kin}

\begin{figure} [!htb]
\centering
\includegraphics[scale=1.39]{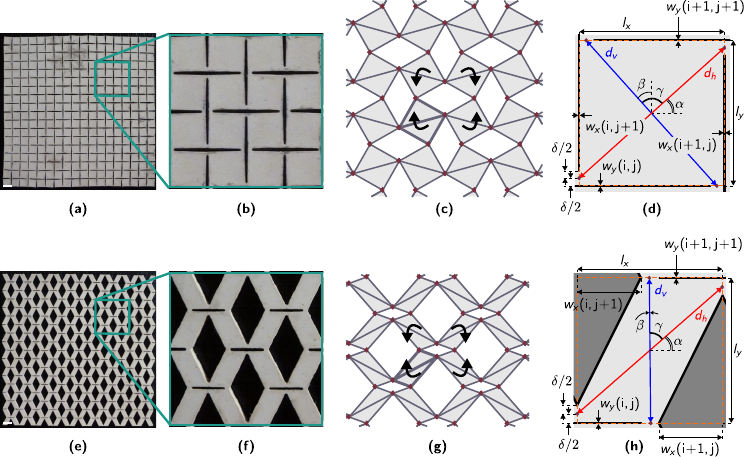}
  \caption{Kinematic analysis of periodic perforated sheets featuring tiles connected by thin hinges. (a) Isotropic sheet and (b) detail. (c) Mechanism of inextensional deformation for the truss analog of (a), obtained by computing the null space of the kinematic matrix. (d) Detail of one of the tiles of (a), indicating all the quantities necessary for the kinematic analysis. (e-h) Same as (a-d), but for the architecture in (e). (Scale bar, 6 mm)}
\label{fig:mechanism}
\end{figure}

The sheets discussed in this work are designed to display mechanisms of inextensional deformation, i.e., low energy modes of compliant mechanism-like deformation. In this Section, we consider the pin-jointed truss analogs of some of our sheets, and resort to the matrix analysis detailed by Pellegrino \& Calladine~\cite{Pellegrino_IJSS_1986} and Hutchinson \& Fleck~\cite{Hutchinson_JMPS_2006} to determine what these mechanisms are. This analysis consists of the following steps. First, we calculate the equilibrium matrix $\textbf{A}$, that relates bar tensions $\textbf{t}$ and joint forces $\textbf{f}$ according to $\textbf{A}\cdot\textbf{t}=\textbf{f}$, and the kinematic matrix $\textbf{B}$, relating joint displacements $\textbf{d}$ and bar elongations $\textbf{e}$ according to $\textbf{B}\cdot\textbf{d}=\textbf{e}$. Note that equilibrium imposes that $\textbf{B}=\textbf{A}^{\mathrm {T}}$. Then, we apply boundary conditions to suppress rigid body motions; in this case, we block the $x$ and $y$ displacements of node (1,1), the node at the bottom left of the specimen, and the $y$ displacement of node (2,1). Finally, we compute the null space of $\textbf{B}$. If the system is properly constrained, each vector belonging to this null space represents a mode of inextensional deformation. The results of this analysis for two cut patterns are shown in Fig.~\ref{fig:mechanism}.
First, we consider the periodic architecture in Fig.~\ref{fig:mechanism}(a-b), known for its auxeticity~{\cite{Grima_PSS_2005}}. The matrix analysis of the pin-jointed truss analog to this system predicts only one mechanism, shown in Fig.~\ref{fig:mechanism}(c), and characterized by the tile rotations highlighted by the black arrows. Note that this geometry features no states of self-stress. Thus, even though the analysis assumes small deformations, the same mechanism should extend to large stretch regimes~{\cite{Hutchinson_JMPS_2006}}. The periodic sheet in Fig.~\ref{fig:mechanism}(e) (same as the one shown in Fig.~\ref{fig:tensile}(a)) features a very similar mechanism of inextensional deformation, characterized by the same relative rotations of the tiles, but with an equivalent positive Poisson's ratio. The tensile tests in Fig.~\ref{fig:tensile}(a) demonstrate that the rubber sheets, despite presenting non-idealities such as finite-sized hinges, deform according to the corresponding mechanisms up to certain stretch values.

Knowing how these periodic sheets deform in plane, we resort to a kinematic model in order to quantify their mechanism-like deformation. The unit cells for these periodic architectures consist of four adjacent tiles. It is sufficient to consider a single tile to determine the whole system's response. In Fig.~\ref{fig:mechanism}(d,h) we show a single tile from the sheets in Fig.~\ref{fig:mechanism}(a,e), such that $(i + j) / 2 \in \mathbb{Z}$, and we indicate all the useful geometric parameters. Here, $(i, \hspace{0.17em} j)$ indicates a generic tile, with $i = 1, ..., N_x + 1$, $j = 1, ..., N_y + 1$ and $N_x$, $N_y$ being the number of tiles along the horizontal and vertical directions. Note that, if we consider a tile such that $(i + j) / 2 \not\in \mathbb{Z}$, the following formulae will only slightly vary. For the remainder of this section, we assume that we are dealing with periodic architectures; this implies that $w_y (i, j) = w_y  (i + 1, j + 1) = w_y$, and $w_x  (i + 1, j) = w_x  (i, j + 1) = w_x$. The red and blue lines indicate the diagonals of each tile. Their lengths are
\begin{equation}
d_h=\sqrt{l_x^2+[l_y-2w_y-\delta]^2}\,\,\,\,\,\,\mathrm{and}\,\,\,\,\,\,d_v=\sqrt{l_y^2+[l_x-2w_x-\delta]^2}\,.
\end{equation}
Ideally, tiles can rotate until the diagonal lines corresponding to the selected stretch direction are straightened. With this in mind, we can determine the maximum horizontal and vertical stretches for any periodic architecture designed following our paradigm, as
\begin{equation}
\lambda_x^M=\frac{d_h}{l_x}\,\,\,\,\,\,\mathrm{and}\,\,\,\,\,\,\lambda_y^M=\frac{d_v}{l_y}\,.
\label{eq:maxstretch}
\end{equation}
We can also use kinematics to derive formulae for the tangential stretches as functions of $\lambda_x^M$ or $\lambda_y^M$. First, we determine the angle $\alpha$ between $d_h$ and the $x$-axis in the undeformed configuration, and $\beta$ between $d_v$ and the $y$-axis, as
\begin{equation}
\alpha=\arctan\left(\frac{l_y-2w_y-\delta}{l_x}\right)\,\,\,\,\,\,\mathrm{and}\,\,\,\,\,\,\beta=\arctan\left(\frac{2w_x+\delta-l_x}{l_y}\right)\,.
\end{equation}
Note that we define $\alpha$ to be positive counterclockwise and $\beta$ to be positive clockwise. We also define $\gamma=\pi/2-\alpha-\beta$ as the angle between $d_h$ and $d_v$. During mechanism-like deformation, $\gamma$ remains fixed since we assume the tiles are rotating rigidly. On the other hand, the inclinations of $d_h$ and $d_v$ with respect to $x$ and $y$ change during the deformation process. To determine the intermediate stages of the sheet's deformation, we define $\alpha^*$ and $\beta^*$ as angles varying from 0 to $\alpha$ and 0 to $\beta$, respectively. Consider now the case of stretching along $x$. We can write
\begin{equation}
\lambda_x(\alpha^*)=\frac{d_h\cos\alpha^*}{l_x}\,\,\,\,\,\,\mathrm{and}\,\,\,\,\,\,\lambda_y(\alpha^*)=\frac{d_v\sin(\gamma+\alpha^*)}{l_y}\,.
\end{equation}
From the first of the two equations, we obtain $\alpha^*(\lambda_x)$ as
\begin{equation}
\alpha^*(\lambda_x)=\arccos\left( \frac{\lambda_x l_x}{d_h} \right)\,.
\end{equation}
Substitution leads to the following formula for $\lambda_y(\lambda_x)$:
\begin{equation}
\lambda_y(\lambda_x)=\frac{d_v}{l_y}\sin\left[\gamma+\arccos\left( \frac{\lambda_x l_x}{d_h} \right)\right]\,.
\end{equation}
This formula is used to determine the analytical curves in Fig.~\ref{fig:tensile}(f), representing the evolution of the tangential stretch as a function of the applied one. Note that a similar formula can be obtained for $\lambda_x(\lambda_y)$.

\begin{figure} [!htb]
\centering
\includegraphics[scale=1.40]{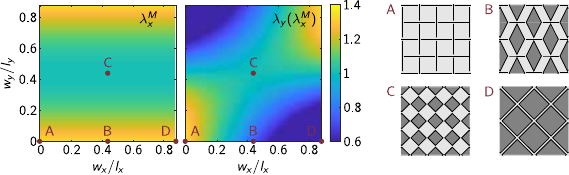}
  \caption{ Design space in terms of maximum stretches, $\lambda_x^M$ and $\lambda_y(\lambda_x^M)$, as a function of $w_x$ and $w_y$, with $l_x=l_y=6\,\mathrm{mm}$ and $\delta=l_x/8$ fixed. Insets A-D represent specific examples extracted from the space.}
\label{fig:designspace}
\end{figure}

In our work, we fix the design parameters $l_x$, $l_y$ and $\delta$ most of the time, and vary $w_x$ and $w_y$. Different combinations of $w_x$ and $w_y$ allow to span a wide design space in terms of achievable deformations. To get a better idea of the available design space, in Fig.~\ref{fig:designspace}, we report plots for the maximum stretch $\lambda_x^M$, and the related
tangential stretch $\lambda_y (\lambda_x^M)$, as a function of $w_x$ and
$w_y$.
Note that the values in the colormaps are specific for $l_x=l_y=6\,\mathrm{mm}$ and $\delta=l_x/8$. We can see that choosing $w_x$ and $w_y$ allows to obtain a wide range of responses to stretching. Some significant examples (A, B, C and D) are extracted from the design space. A, corresponding to $w_x=w_y=0$, is characterized by $\lambda_x^M=\lambda_y(\lambda_x^M)=1.33$; B, corresponding to $w_x=(l_x-\delta)/2$ and $w_y=0$, is characterized by $\lambda_x^M=1.33$ and $\lambda_y(\lambda_x^M)=0.75$; C, corresponding to $w_x=(l_x-\delta)/2$ and $w_y=(l_y-\delta)/2$, is kinematically undeformable albeit featuring bulky tiles connected by thin hinges; D, corresponding to $w_x=l_x-\delta$ and $w_y=0$, does not behave like a mechanism since the rigid tiles assumption does not hold for these specific parameters. From these examples, it is clear that not all the configurations available in the design space allow to obtain the in-plane mechanism-like deformation behavior we are interested in. Therefore, particular care is needed when choosing the design parameters; in light of this, in this work, we limit ourselves to the ranges $0 \leq w_x \leq (l_x-\delta)/2$ and $0 \leq w_y \leq (l_y-\delta)/2$.

\begin{figure} [!htb]
\centering
\includegraphics[scale=1.39]{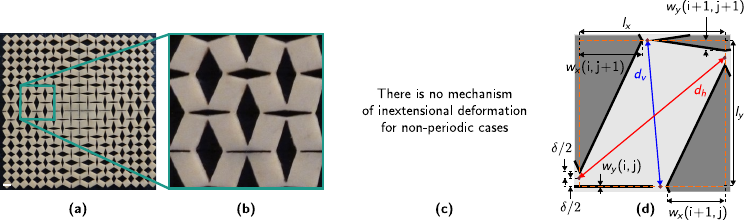}
  \caption{Kinematic analysis of non-periodic perforated sheets featuring tiles connected by thin hinges. (a) Example of non-periodic sheet. (b) Detail of the sheet in (a). (c) The null space of the kinematic matrix of the pin-jointed truss analog to (a) contains no mechanism. (d) Detail of one of the tiles of the sheet in (a), with all the quantities necessary for the kinematic analysis. (e-h) Same as (a-d), but for the architecture in (e). (Scale bar, 6 mm)}
\label{fig:nonmechanism}
\end{figure}

An example of non-periodic sheet is shown in Fig.~\ref{fig:nonmechanism}(a). 
Non-periodicity leads to frustration and to the disappearance of mechanisms of inextensional deformation. This is confirmed by the matrix analysis of the pin-jointed truss analog of the architecture in Fig.~\ref{fig:nonmechanism}(a), that has no mechanisms. In these non-periodic cases, we can still use kinematics to infer something about the local deformation of the sheet, even though it cannot be used to quantify global deformations as it did in periodic scenarios. For this reason, in the main article, we sometime consider the maximum stretches that a tile belonging to a non-periodic sheet can undergo. We interpret these stretches as measures of a local ability to deform. The local ability to behave like a mechanism is what makes these systems buckle out of plane. For a generic tile in a non-periodic scenario, whose bottom-left gridpoint $(i,j)$ is such that $(i+ j)/2\in \mathbb{Z}$, the maximum stretches are calculated as in Eq.~\ref{eq:maxstretch}, with $d_h$ and $d_v$ computed as
\begin{equation}
d_h=\sqrt{l_x^2+[l_y-w_y(i,j)-w_y(i+1,j+1)-\delta]^2}\,\,\,\,\,\,\mathrm{and}\,\,\,\,\,\,d_v=\sqrt{l_y^2+[l_x-w_x(i,j+1)-w_x(i+1,j)-\delta]^2}\,.
\end{equation}

\clearpage
\section{Cut pattern generation}
\label{sec:patt}
Our cut patterns are generated and kinematically analyzed using custom MATLAB scripts. The first step of the design process is to generate a grid of points. The grid can be non-Cartesian, as long as it can be mapped to a rectangular one. At each grid point $(i,\,j)$ with $i=1,...,N_x+1$ and $j=1,...,N_y+1$ with $N_x$ and $N_y$ being the number of tiles along the horizontal and vertical directions, we cut a diamond-shaped hole. For each diamond, we define either its horizontal or vertical half-diagonal, i.e. $w_x$ or $w_y$. If $(i+ j)/2\in \mathbb{Z}$ we define the diamond's $y$-oriented half-diagonal $w_y(i,\,j)$. Its $x$-dimension will be determined by the neighboring diamonds---$l_x-\delta-w_x(i-1,j)$ to the left and $l_x-\delta-w_x(i+1,j)$ to the right of the grid point. Otherwise, if $(i+ j)/2\not\in \mathbb{Z}$, we define $w_x(i,\,j)$ while the diamond's $y$-dimension follows from the neighboring diamonds. This design paradigm guarantees geometric continuity and that no perforations overlap, even in non-periodic architectures where we let $w_x$ and $w_y$ vary (smoothly or not) from diamond to diamond. In the case of architectures designed to allow for plastic deformations, instead of defining a diamond, we define an octahedron at each gridpoint.

The $w_x$, $w_y$ functions corresponding to all cut patterns shown throughout this manuscript are listed in the following.
\begin{itemize}
  \item ``Anisotropic'' sheet.\\
  Appearing in Fig.~\ref{fig:tensile}(a,d), Fig.~\ref{fig:mechanism}(e), Fig.~\ref{fig:fem}(a), Fig.~\ref{fig:designspace}B, Fig.~\ref{fig:tensile2}, Fig.~\ref{fig:anisotropicmicro}, Fig.~\ref{fig:tensile4}.\\
  Loading: Uniform horizontal or uniform vertical.\\
  Material: Natural rubber gum of various thicknesses ($1.55$, $3.1\,$ mm).\\
  Parameters: $N_x=N_y=18$, $l_x=l_y=6\,$mm, $\delta=l_x/8$.\\
  Hole size distribution (with $i=1, ..., N_x+1$, $j=1, ..., N_y+1$):
  \begin{equation*}
  w_x(i,j)=(l_x-\delta)/2\,\,,\,\,\,\,\,\,
  w_y(i,j)=0\,\,.
  \end{equation*}

  \item ``Isotropic'' sheet.\\
  Appearing in Fig.~\ref{fig:tensile}(c), Fig.~\ref{fig:mechanism}(a), Fig.~\ref{fig:designspace}A, Fig.~\ref{fig:tensile3}.\\
  Loading: Uniform horizontal or uniform vertical.\\
  Material: Natural rubber gum, $1.55\,$mm thick.\\
  Parameters: $N_x=N_y=18$, $l_x=l_y=6\,$mm, $\delta=l_x/8$.\\
  Hole size distribution (with $i=1, ..., N_x+1$, $j=1, ..., N_y+1$):
  \begin{equation*}
  w_x(i,j)=0\,\,,\,\,\,\,\,\,
  w_y(i,j)=0\,\,.
  \end{equation*}

  \item ``Unstretchable'' sheet.\\
  Appearing in Fig.~\ref{fig:tensile}(e), Fig.~\ref{fig:designspace}C.\\
  Loading: Uniform horizontal or uniform vertical.\\
  Material: Natural rubber gum, $1.55\,$mm thick.\\
  Parameters: $N_x=N_y=18$, $l_x=l_y=6\,$mm, $\delta=l_x/8$.\\
  Hole size distribution (with $i=1, ..., N_x+1$, $j=1, ..., N_y+1$):
  \begin{equation*}
  w_x(i,j)=(l_x-\delta)/2\,\,,\,\,\,\,\,\,
  w_y(i,j)=(l_y-\delta)/2\,\,.
  \end{equation*}

  \item ``Graded'' or ``Dome'' sheet.\\
  Appearing in Fig.~\ref{fig:graded}, Fig.~\ref{fig:gradedparameters}, Fig.~\ref{fig:gradedparameters2}, Fig.~\ref{fig:plasticitydome}(a).\\
  Loading: Horizontal point loads at $y=y^M/2$ along the left and right boundaries.\\
  Material: Natural rubber gum of various thicknesses ($1.55$, $3.1\,$ and $0.75\,$ mm).\\
  Parameters: $N_x=36$, $N_y=18$, $l_x=6\,$mm, $l_y=2l_x$, $\delta=l_x/8$.\\
  Hole size distribution (with $i=1, ..., N_x+1$, $j=1, ..., N_y+1$):
  \begin{equation*}
  w_x(i,j)=\frac{(l_x-\delta)}{2}\,\,,\,\,\,\,\,\,
  w_y(i,j)=\frac{l_y-\delta}{2} \left|\cos\frac{j\pi}{N_y+2}\right|\,\,.
  \end{equation*}

  \item ``Two bumps'' sheet.\\
  Appearing in Fig.~\ref{fig:nonperiodic}(a1-a2) and Fig.~\ref{fig:fem}(b).\\
  Loading: Point loads at the four corners, directed along $\pm 5^{\mathrm{o}}$ with respect to the horizontal.\\
  Material: Natural rubber gum, $1.55\,$mm thick.\\
  Parameters: $N_x=37$, $N_y=18$, $l_x=l_y=6\,$mm, $\delta=l_x/8$.\\
  Hole size distribution (with $i=1, ..., N_x+1$, $j=1, ..., N_y+1$):
  \begin{equation*}
  w_x(i,j)=\left \{\!\!\!\begin{array}{l}
    \frac{l_x-\delta}{2} \left|\cos\frac{i\pi}{(N_x+1)/2}\right|\,\,\,\,\,\,\mathrm{if}\,\,i<(N_x+1)/2+1\\
     \\
    \frac{l_x-\delta}{2} \left|\cos\frac{(i-(N_x+1)/2)\pi}{(N_x+1)/2}\right|\,\,\,\,\,\,\mathrm{if}\,\,i \geq (N_x+1)/2+1 \end{array}\right.
    \,\,,\,\,\,\,\,\,
  w_y(i,j)=\frac{l_y-\delta}{2} \left|\cos\frac{j\pi}{N_y+2}\right|\,\,.
  \end{equation*}

  \item ``Flower'' sheet.\\
  Appearing in Fig.~\ref{fig:nonperiodic}(b1-b3).\\
  Loading: Point loads at the four corners (directed at $\pm 45^{\mathrm{o}}$ with respect to the horizontal), or point loads at the centerpoints of the four edges (and perpendicular to those edges).\\
  Material: Natural rubber gum, $1.55\,$mm thick.\\
  Parameters: $N_x=37$, $N_y=37$, $l_x=l_y=6\,$mm, $\delta=l_x/8$.\\
  Hole size distribution (with $i=1, ..., N_x+1$, $j=1, ..., N_y+1$):
  \begin{equation*}
  w_x(i,j)=\frac{l_x-\delta}{2} \left|\cos\frac{2i\pi}{N_x+1}\right| \left|\cos\frac{2j\pi}{N_y+1}\right|\,\,,
  \,\,\,\,\,\,
  w_y(i,j)=\frac{l_y-\delta}{2} \left|\cos\frac{2i\pi}{N_x+1}\right| \left|\cos\frac{2j\pi}{N_y+1}\right|\,\,.
  \end{equation*}

  \item ``C pattern'' sheet.\\
  Appearing in Fig.~\ref{fig:nonperiodic}(c1-c2).\\
  Loading: Point loads at few points along each boundary. All loads are perpendicular to the boundaries.\\
  Material: Natural rubber gum, $1.55\,$mm thick.\\
  Parameters: $N_x=30$, $N_y=30$, $l_x=l_y=6\,$mm, $\delta=l_x/8$.\\
  Hole size distribution: We did not use analytical functions of $i$ and $j$ to create this pattern. The $w_x$, $w_y$ couples we use are $w_x=(l_x-\delta)/2$ and $w_y=(l_y-\delta)/2$ outside the C, and $w_x=0$, $w_y0$ inside the C.

  \item ``Bulging tube''.\\
  Appearing in Fig.~\ref{fig:last}(a).\\
  Loading: Axial loads applied at the ends of the tube through 3D printed rings.\\
  Material: Natural rubber gum, $1.55\,$mm thick.\\
  Parameters: $N_x=18$, $N_y=40$, $l_x=l_y=6\,$mm, $\delta=l_x/8$.\\
  Hole size distribution (with $i=1, ..., N_x+1$, $j=1, ..., N_y+1$):
  \begin{equation*}
  w_x(i,j)=0
    \,\,,\,\,\,\,\,\,
  w_y(i,j)=\left \{\!\!\!\begin{array}{l}
    (l_y-\delta)/2 \,\,\,\,\,\,\mathrm{if}\,\,j\leq6\,|\,(j\geq15\,\&\,j\leq20)\,|\,(j\geq26\,\&\,j\leq31)\,|\,j\geq35\\
     \\
  0 \,\,\,\,\,\,\mathrm{if}\,\,(j\geq7\,\&\,j\leq14)\,|\,(j\geq21\,\&\,j\leq26)\,|\,(j\geq32\,\&\,j\leq34)\end{array}\right.
  \,\,.
  \end{equation*}

  \item ``Petal'' sheet.\\
  Appearing in Fig.~\ref{fig:last}(b).\\
  Loading: Axial loads applied at the petal's extremities.\\
  Material: Natural rubber gum, $1.55\,$mm thick.\\
  Parameters: We used $w_x=(l_x-\delta)/2$ and $w_y=(l_y-\delta)/2$ along the petal's boundaries and in those regions that we want to remain stiff; we used $w_x=0$, $w_y=0$ elsewhere.


  \item ``Plastic vase''.\\
  Appearing in Fig.~\ref{fig:last}(c).\\
  Loading: Manual forming.\\
  Material: $0.5\,$mm-thick PETG sheet.\\
  Parameters: $N_x=36$, $N_y=18$, $l_x=6\,$mm, $l_y$ varies linearly from $6\,$mm at the bottom of the specimen to $18\,$mm at the top, $\delta=l_x/10$, $h=l_x/8$.\\
  Hole size distribution (with $i=1, ..., N_x+1$, $j=1, ..., N_y+1$):
  \begin{equation*}
  w_x(i,j)=\frac{(l_x-\delta)}{2}\,\,,\,\,\,\,\,\,
  w_y(i,j)=\frac{l_y-\delta}{2}-\frac{l_y-\delta}{2} \left|\cos\frac{j\pi}{2N_y+4}\right|\,\,.
  \end{equation*}

  \item ``No mechanism'' sheet.\\
  Appearing in Fig.~\ref{fig:nonmechanism}(a).\\
  Loading: None.\\
  Material: Natural rubber gum, $1.55\,$mm thick.\\
  Parameters: $N_x=18$, $N_y=18$, $l_x=l_y=6\,$mm, $\delta=l_x/8$.\\
  Hole size distribution (with $i=1, ..., N_x+1$, $j=1, ..., N_y+1$):
  \begin{equation*}
  w_x(i,j)=\frac{l_x-\delta}{2} \left|\cos\frac{i\pi}{N_x+2}\right|
    \,\,,\,\,\,\,\,\,
  w_y(i,j)=\frac{l_y-\delta}{2} \left|\cos\frac{j\pi}{N_y+2}\right|\,\,.
  \end{equation*}

  \item ``Plastic dome''.\\
  Appearing in Fig.~\ref{fig:plasticitydome}(b).\\
  Loading: Horizontal point loads at $y=y^M/2$ along the left and right boundaries.\\
  Material: $0.5\,$mm-thick PETG sheet.\\
  Parameters: $N_x=36$, $N_y=18$, $l_x=6\,$mm, $l_y=2l_x$, $\delta=l_x/10$, $h=l_x/8$.\\
  Hole size distribution (with $i=1, ..., N_x+1$, $j=1, ..., N_y+1$):
  \begin{equation*}
  w_x(i,j)=\frac{(l_x-\delta)}{2}\,\,,\,\,\,\,\,\,
  w_y(i,j)=\frac{l_y-\delta}{2} \left|\cos\frac{j\pi}{N_y+2}\right|\,\,.
  \end{equation*}

\end{itemize}

\clearpage
\section{Influence of the design parameters on the in-plane deformation of periodic specimens}
\label{sec:par}
\begin{figure*} [!htb]
\centering
\includegraphics[scale=1.40]{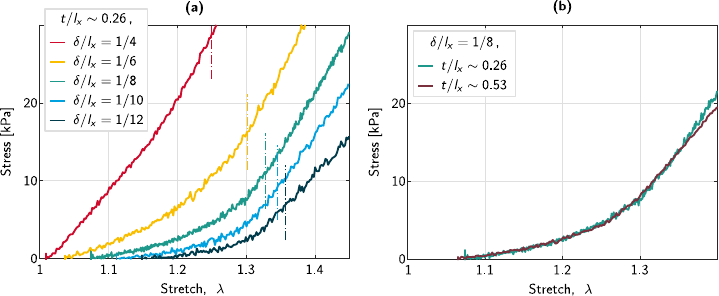}
	\caption{ (a) Dependence of the mechanism-like response on the in-plane hinge width, $\delta$. We keep $t / l_x \sim 0.26$ constant, and we vary $\delta / l_x$. The dashed vertical lines represent the mechanism-to-elasticity transitions for all $\delta / l_x$ cases. (b) Dependence of the  mechanism-like response on the sheet's thickness, $t$, with $\delta / l_x = 1 / 8$ constant.}
\label{fig:tensile4}
\end{figure*}

To analyze the influence of the design parameters on the in-plane response of perforated sheets, we consider the cut pattern discussed in Fig.~\ref{fig:tensile} as reference case. The results of this analysis are reported in Fig.~\ref{fig:tensile4}.
In Fig.~\ref{fig:tensile4}(a), we show the dependence of the horizontal stretch response on the in-plane width of the hinges $\delta$, for a constant out-of-plane thickness of the sheet $t/l_x\sim 0.26$ (corresponding to $t=1.55\,\mathrm{mm}$). If $\delta$ is increased, the sheet tends to lose its mechanism-like behavior. This is evident from the fact that the red and yellow continuous curves do not display a clear mechanism-to-elasticity transition. On the other hand, this transition is more pronounced for small $\delta$. Note that the dash-dot lines represent the mechanism-to-elasticity transitions for each $\delta$ value. They are different from each other since the lengths of the tile diagonals $d_h$ and $d_v$ differ when we change $\delta$.

In Fig.~\ref{fig:tensile4}(b), we superimpose the responses of two specimens featuring the same architecture with $\delta / l_x = 1 / 8$, and different sheet thicknesses, $t$. We observe that the two responses overlap in the mechanism region, and in part of the elasticity-dominated regime. The curves deviate for stretches larger than 1.35.

\clearpage
\section{Influence of the design parameters on the out-of-plane deformation of non-periodic sheets}
\label{sec:gpar}
\begin{figure*} [!htb]
\centering
\includegraphics[scale=1.40]{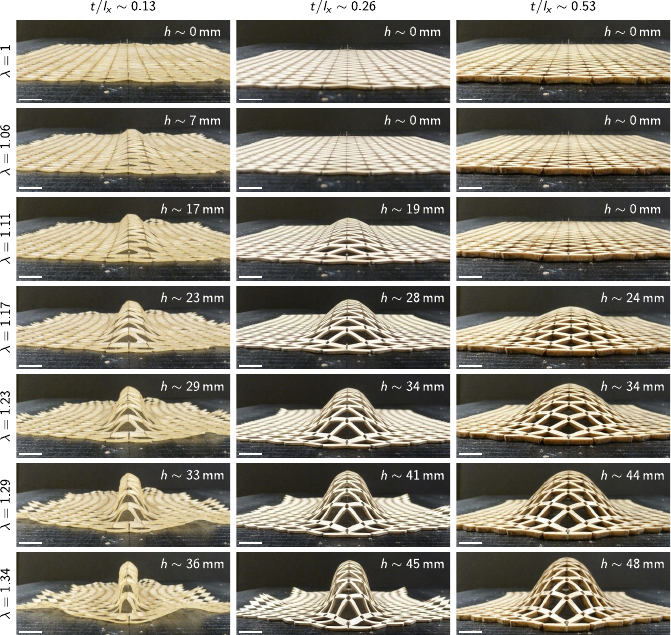}
  \caption{ Out-of-plane deformation of three graded sheets with different thicknesses, for different stretches. Rows of images correspond to specific stretch values. Columns correspond to different thicknesses of the sheets. In each image, $h$ indicates the height of the highest point of the 3D shape with respect to its undeformed position. (Scale bar, 12 mm)}
\label{fig:gradedparameters}
\end{figure*}

\begin{figure*} [!htb]
\centering
\includegraphics[scale=1.40]{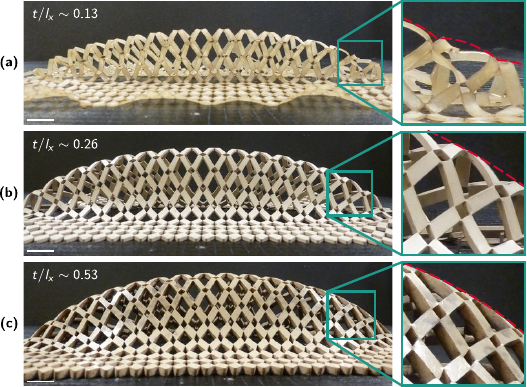}
  \caption{Out-of-plane deformation of three graded sheets with different thicknesses, for the same stretch value. (a) Corresponds to $t/l_x\sim 0.13$, (b) to $t/l_x\sim 0.26$ and (c) to $t/l_x\sim 0.53$. The left images represent lateral views of the buckled shapes. The details highlight the local deformations of hinges and tiles near the load application points. (Scale bar, 12 mm)}
\label{fig:gradedparameters2}
\end{figure*}

Fig.~\ref{fig:gradedparameters} and Fig.~\ref{fig:gradedparameters2} provide information on the influence of $\delta$ and $t$ on the doming of an elastic sheet. This information is also reported in a concise way in Fig.~\ref{fig:graded}(c,d). Note that these shapes have been obtained by 1) pulling the specimen by hand up to a desired stretch value, 2) nailing it to a wooden board, 3) pinching the center of the specimen to trigger out-of-plane buckling. This guarantees that all the images in Fig.~\ref{fig:gradedparameters} and Fig.~\ref{fig:gradedparameters2} are obtained with consistent loading conditions. It also ensures that, if two stable solutions exist for a certain stretch value, we jump on the one that corresponds to out-of-plane deformation. For these reasons, the critical buckling stretches observed in experiments made with tensile test apparati are bound to differ from the results shown here.

In addition to the comments in the main text, we here discuss the influence of $\delta/t$. When $t$ is decreased below the in-plane hinge width $\delta$, the out-of-plane (rather than the in-plane) bending of the hinges becomes favorable: this translates into the formation of localized crease patterns~{\cite{Rafsanjani_PRL_2017,Rafsanjani_SCIROB_2018}}. In our case, this behavior introduces local undulations superimposed to the global three-dimensional profile and concentrated near the loading sites. This is shown in Fig.~\ref{fig:gradedparameters2}(c).
%


\clearpage
\section{Localized vs. distributed loads}
\label{sec:distr}
\begin{figure*} [!htb]
\centering
\includegraphics[scale=1.40]{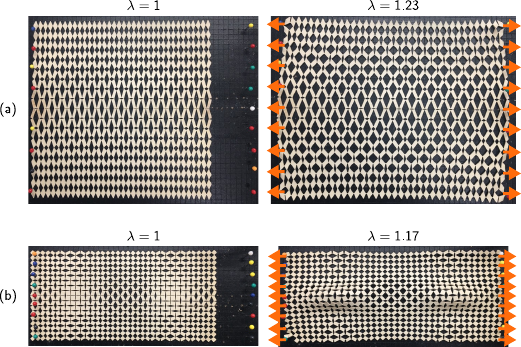}
    \caption{(a) Deformation of the graded specimen of Fig.~\ref{fig:graded}, subjected to a distributed horizontal stretch. (b) Deformation of the specimen introduced in Fig.~\ref{fig:nonperiodic}a1-a2, subjected to a uniform horizontal stretch. The direction and points of application of the excitation are marked with orange arrows.}
\label{fig:distrib}
\end{figure*}

All non-periodic specimens analyzed in our article have been loaded via localized boundary excitations. In this Section, we discuss the response of some of those specimens to distributed boundary loads. In Fig.~\ref{fig:distrib}(a), we can see the unstretched graded specimen (introduced in Fig.~\ref{fig:graded}), and its deformed configuration when subjected to a uniform stretch $\lambda=1.23$. 
Clearly, distributed loads do not produce any out-of-plane buckling. This is not surprising, since this loading scenario produces a uniform stretch of each row of identical unit cells, with the central rows featuring lower stresses than the top and bottom boundary regions. In this case, no significant compressive stresses are triggered and no geometric frustration arises. In Fig.~\ref{fig:distrib}(b), we show the response of the specimen with two auxetic islands (introduced in Fig.~\ref{fig:nonperiodic}(a1-a2) to a similar uniform load. In this case, we stretch the specimen to $\lambda=1.17$. We can see that out-of-plane buckling occurs and that the two auxetic regions pop-up, just like in the point-loading scenario. This can be ascribed to the fact that buckling-inducing compressive stresses can still arise due to uniform loading, owing to the presence of auxetic islands surrounded by non-auxetic regions. We also notice that the buckled pattern obtained with distributed loads features bumps that are more elongated in the horizontal direction; this further proves that the loading configuration can be leveraged to control the shape of the buckled features.

\clearpage
\section{Alternative design for stiff materials and plastic deformations}
\label{sec:stiff}

\begin{figure} [!htb]
\centering
\includegraphics[scale=1.40]{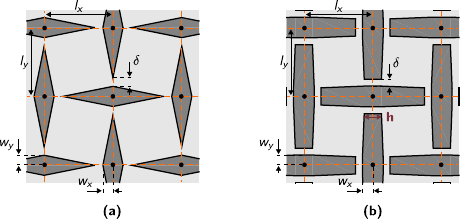}
  \caption{ An alternative design startegy for stiff materials. (a) Detail of an architecture obtained with our initial design strategy. (b) Compliant beam version of the same architecture, where we have introduced the additional parameter $h$, representing the in-plane hinge length.}
\label{fig:plasticityuc}
\end{figure}

In order to fabricate sheets out of stiff materials, and to have our sheets retain their 3D shape upon load removal, we slightly modify our cut pattern design. To achieve shape retention, we leverage plastic deformations that occur at the hinges when elastic-plastic materials are used. If the same hinge geometry used for soft materials were used for stiff ones, both periodic and non-periodic specimens would shatter at the hinges when pulled open. This is why we modify our hinge design. To do so, we follow the guidelines offered by Shang, Pasini et al.~{\cite{Shang_JMR_2018}}. This entails defining octahedra-shaped cuts instead of diamond-shaped ones at each grid point. This new design is illustrated in Fig.~\ref{fig:plasticityuc}(b). It represents the compliant mechanism version of the architecture in Fig.~\ref{fig:plasticityuc}(a).
As a result, the hinges produced with the new design have a finite length $h$. Note that the overall response of this alternate geometry is similar to the original one. The requirement is for the hinge length $h$ to be much smaller than the distances between gridpoints, $l_x$ and $l_y$.

\begin{figure*} [!htb]
\centering
\includegraphics[scale=1.40]{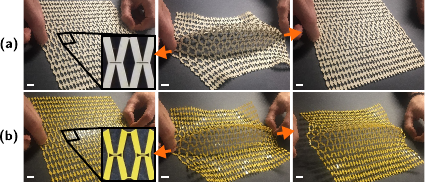}
  \caption{ (a) Three stages of the deformation of the natural rubber specimen studied in Fig.~3(a). (b) Deformation of a similar sheet, made of PETG and featuring the design variation shown in Fig.~\ref{fig:plasticityuc}(b). (Scale bar, 12 mm)}
\label{fig:plasticitydome}
\end{figure*}

In Fig.~\ref{fig:plasticitydome}, we compare the response of the natural rubber sheet also shown in Fig.~3(a1), to the response of a sheet made of PETG, featuring a similar cut pattern albeit modified by selecting $\delta = l_x / 10$ and introducing $h = l_x / 8$. Upon load removal, the PETG sheet partially retains its deformed, three-dimensional shape, while the rubber one does not.

\end{document}